\documentclass[a4paper]{article}

\usepackage{INTERSPEECH2022}

\usepackage{soul}
\usepackage{xcolor}
\usepackage{booktabs}
\usepackage{multicol}
\usepackage{multirow}
\usepackage{graphicx}
\usepackage{markdown}
\usepackage{subcaption}

\newcommand{\citem}[1]{
  \IfEq{#1}{}
      {\sethlcolor{orange}\hl{[?]}\sethlcolor{yellow}}
      {\cite{#1}}
}
\newcommand{\myparagraph}[1]{
  \vspace{2pt}
  \noindent \textbf{#1:} 
}

\AtBeginDocument{
  \setlength\abovedisplayskip{0pt}
  \setlength\belowdisplayskip{0pt}}
\usepackage[margin=0pt, font=small]{caption}
\title{Predicting score distribution to improve non-intrusive speech quality estimation\thanks{Submitted to Interspeech 2022}}
\name{Abu Zaher Md Faridee$^{1\ast}$\thanks{$^\ast$Work performed while intern at Microsoft Research.}, Hannes Gamper$^2$}
\address{
  $^1$Information Systems, University of Maryland, Baltimore County, MD, USA\\
  $^2$Microsoft Research Labs, Redmond, WA, USA}
\email{faridee1@umbc.edu, hagamper@microsoft.com}

\begin{document}

\maketitle

\begin{abstract}
Deep noise suppressors (DNS) have become an attractive solution to remove background noise, reverberation, and distortions from speech and are widely used in telephony/voice applications. 
They are also occasionally prone to introducing artifacts and lowering the perceptual quality of the speech.
Subjective listening tests that use multiple human judges to derive a mean opinion score (MOS) are a popular way to measure these models' performance.
Deep neural network based non-intrusive MOS estimation models have recently emerged as a popular cost-efficient alternative to these tests. 
These models are trained with only the MOS labels, often discarding the secondary statistics of the opinion scores.
In this paper, we investigate several ways to integrate the distribution of opinion scores (e.g. variance, histogram information) to improve the MOS estimation performance. 
Our model is trained on a corpus of 419K denoised samples  by 320 different DNS models and model variations and evaluated on 18K test samples from DNSMOS. 
We show that with very minor modification of a single task MOS estimation pipeline, these freely available labels can
provide up to a  0.016 RMSE and 1\% SRCC improvement.
\end{abstract}
\vspace{2pt}

\section{Introduction}

As more people are increasingly working from home and using live telephony and communication applications to collaborate with their peers as well as stay connected to friends and family, retaining and improving speech quality has become a topic of immense importance in industry and academia~\cite{dubey2022icassp,reddy2021icassp,reddy2020interspeech,reddy2021interspeech}. 

Real-time speech enhancement (SE) solutions~\cite{ephraim1984speech,reddy2017individualized} have traditionally been used for decades to improve the perceptual quality of speech. Nowadays they are being replaced by Deep Noise Suppression (DNS)~\cite{fu2017raw,choi2020phase,koyama2020exploring} models due to their flexibility in handling a variety of background noises, room reverberations, and distortions. 
However, due to the possible wide variety in the training datasets and model architecture, each DNS model often performs noticeably better and worse in dealing with certain kinds of noise compared to other models. Moreover, they can also introduce their own set of artifacts
-- ranging from mistaking actual speech for noise and removing it to introducing distortions during the speech reconstruction phase -- all of which can lower the perceptual quality of the speech to the point that an independent listener might prefer the original version of the speech vs the noise suppressed one.

In order to properly provision these DNS models for widespread deployment, their performance needs to be evaluated on a large number of noisy and distorted speech samples. The subjective listening test has been the staple for evaluating the perceived speech signal quality~\cite{itu-t-recommendation-p-800} where multiple users provide judgment on a scale ranging from $1$ to $5$ and usually the average score of all participants over specific condition (commonly referred to as MOS, i.e., mean opinion score) represents the perceived quality after leveling out individual factors~\cite{moller2011speech}. But given the wide number of possible DNS models and noisy sample combinations, they would require huge time and human labor investment and even then cannot achieve real-time feedback~\cite{avila2016performance}, thus making the process unsustainable for conducting large-scale experiments.
Several automated objective instrumental quality measures have been proposed and adopted over the years as an alternative (e.g. PESQ~\cite{itu-t-recommendation-p-862}, POLQA~\cite{itu-t-recommendation-p-863}). However, they were optimized to measure compression artifacts rather than degradation introduced by the noise, reverberation, and speech enhancements. These measures are also limited by their need to have access to the original \textit{clean} signals, making the bulk of them \textit{intrusive} and unable to be applied to the speech captured in the wild.

Several deep-learning based \textit{non-intrusive} speech quality assessment models have been proposed recently that aim to tackle this challenge~\cite{avila2019non,reddy2021dnsmos, mittag21_interspeech}. Most of these models are trained in a supervised way with the aim of minimizing the error between the ground truth MOS scores and the predicted MOS scores. 
Recently, attempts have 
been made to incorporate additional information during model training.
To include the effect of individual judges' bias on the MOS labels, MBNET~\cite{leng2021mbnet} is trained using a multi-task loss with an additional bias term, i.e., the difference between the MOS score and the individual judge score. However, it is not clear how this approach might generalize to  
datasets generated via crowd-sourcing based subjective listening tests~\cite{reddy2021dnsmos} that may include hundreds of judges, who may each provide anywhere from one to hundreds of scores. 
MetricNet~\cite{yu2021metricnet} jointly models MOS estimation with a reconstruction objective of the clean speech signal, to estimate Perceptual Evaluation of Speech Quality (PESQ). The model uses the Wasserstein distance between the ground truth PESQ distribution and the model output as a training objective, where the ground truth distribution is either a simple one-hot vector or a soft target around the true PESQ value.  
It should be noted that PESQ has been shown to correlate poorly with human rating when used for evaluating speech enhancement models~\cite{reddy2021dnsmos}. 

Here, we study incorporating the distribution of scores underlying each MOS label for training a speech quality estimation model geared towards evaluating speech enhancement methods.
We hypothesize that in addition to the first moment (mean) of the subjective listening scores, providing extra supervision concerning the distribution of the scores (e.g. second-moment/variance or histogram information) 
may improve model performance and robustness.
To test our hypothesis, we develop a number of models that incorporate the (a) variance/standard deviation, (b) median (c) histogram bins of the opinion scores ($1-5$ scale) into the primary regression loss calculation logic of MOS estimation by either (a) direct prediction of these statistics, (b) weighting the MOS estimations by these statistics (c) directly predicting the opinion scores themselves. We develop a convolutional LSTM model as the primary \textit{backbone} and run experiments with different loss functions to align the distributions. During our experiments, we found that predicting $5$ opinion scores and then aligning the primary and secondary moments (mean and standard deviation) with the ground truth opinion scores provides the best improvement over vanilla MOS estimation.


\vspace{-1em}
\section{Dataset and score distribution}
\label{sec:dataset}

\begin{figure}[!htb]
    \centering
    \includegraphics[width=\linewidth]{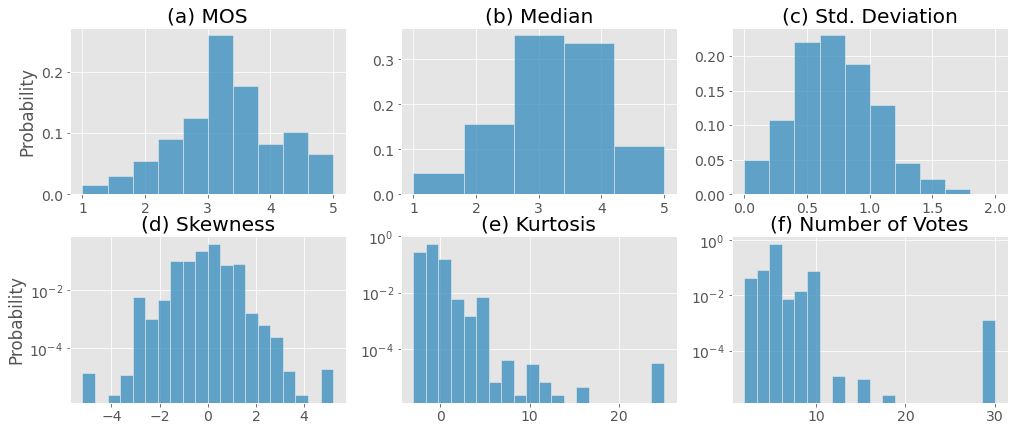}
    \caption{Histogram of (a) MOS, (b) Median, (c) Standard deviation, (d) Skewness, (e) Kurtosis of the scores and (f) Number of opinion scores per clip. The last 3 subfigures are in log-scale for better visibility.}
    \label{fig:hist-labels}
\end{figure}
The dataset used in our experiment is derived from the Interspeech 2020 Deep Noise Suppression Challenge dataset~\cite{reddy2020interspeech}, obtained using ITU-T P.808~\cite{reddy2020interspeech,naderi2020open}. P.808 is an online crowd-sourcing based highly reproducible subjective testing framework. It has been shown to stack rank noise suppression models with high accuracy when each model is tested as an average over a statistically significant number of clips.
In our dataset, 121679 unique files comprising both noisy and clean speech are first processed through 320 unique noise suppression models and model variations. We only take the files that are between 4 and 20 seconds in length and consist of only single-channel 16~kHz samples. The process generates a total of 419836 files in the training set. To allow comparisons with external baselines,
we used the test set from DNSMOS~\cite{reddy2021dnsmos} (18K files) for all evaluations.

The statistics of the training dataset are shown in Figure~\ref{fig:hist-labels}.
The ratings of the speech qualities vary between very poor ($\text{MOS}=1$) and excellent ($\text{MOS}=5$) and as shown in Figures~\ref{fig:hist-labels}(a) and (b), the majority of the MOS ratings are between 2.5 and 4.
From Figure~\ref{fig:hist-labels}(c), we can also see that a sizable number of the samples have opinion scores with a standard deviation, $\sigma > 1$ indicating a high amount of subjectivity in the opinion scores.  
%
The Skewness (Fisher-Pearson) of the opinion scores distribution ranges between -1.75 and 1.75 as shown in Figure~\ref{fig:hist-labels}(d). Such high skewness indicates that the median of the opinion scores 
is often different from the MOS scores.
Interestingly in Figure~\ref{fig:hist-labels}(e), we also notice that majority of the samples are \textit{platykurtic} -- most of the samples are free from extreme outlier opinion scores.
Figure~\ref{fig:hist-labels}(f) demonstrates the number of opinion scores per sample and the majority (75\%) of the samples has 5 opinion scores.


\section{Proposed model architecture}
\label{sec:arch}
\subsection{Backbone Model}
\label{sec:backbone}
\begin{figure}[!htb]
    \centering
    \includegraphics[width=.95\linewidth]{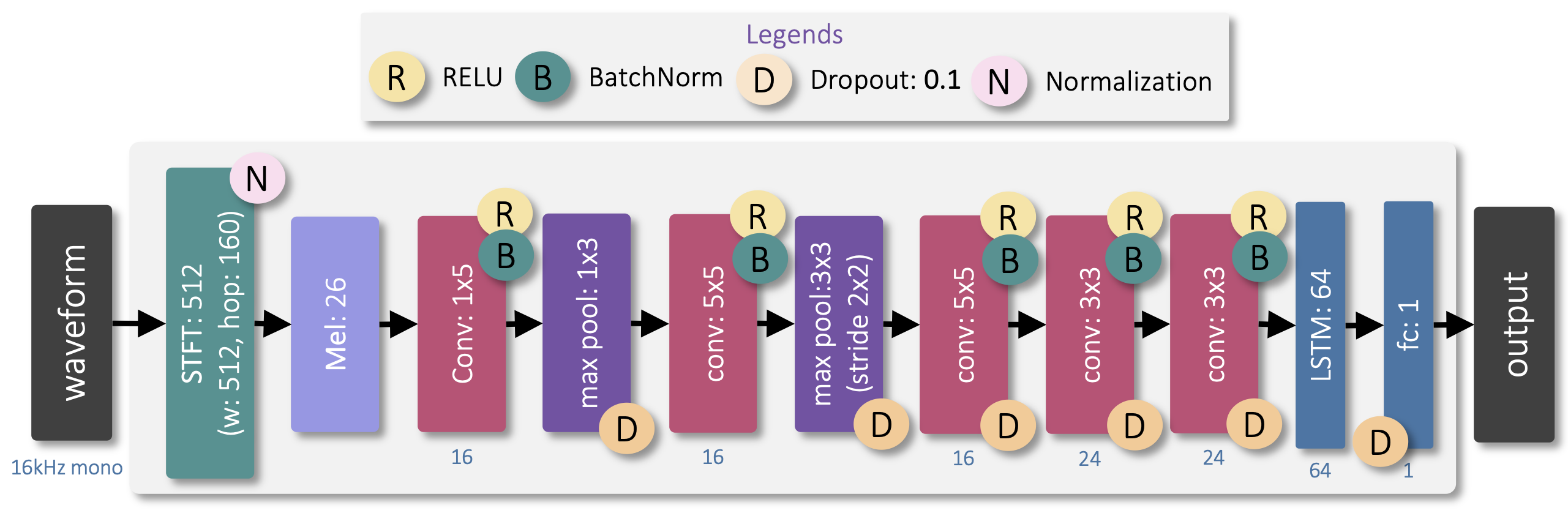}
    \caption{Backbone model (single task MOS estimation) Overview. The shape of the features after each operation is shown at the bottom of each layer.}
    \label{fig:model-overview}
\end{figure}

The $16$~kHz monaural samples are first pre-processed by STFT transform with $512$ samples per frame (i.e., $32$~ms) and a $160$ sample (i.e., $10$~ms) overlap and thereafter $26$ Mel-frequency bins per frame are extracted.
We perform power-to-decibel conversion on the resulting Mel-frequency bins to better align the features with human perception of sound levels.
This results in a $26\times N$ shaped feature matrix per file
where $N$ can be of varying length due to the input audio samples being between $4-20$ seconds long. 
We utilized a convolutional-LSTM based architecture (referred to as \textit{backbone} henceforth) throughout all of our experiments. 
We employ $5$ convolutional layers (without any padding) to gradually decrease the size of the feature space before feeding the resultant features to an LSTM layer. 
The LSTM layer helps to build a fixed-length representation from the variable-length convolutional feature sets. 
The first convolution layer has a $1 \times 5$ shaped kernel 
followed by $1 \times 3$ max-pool operation which helps to capture the temporal relationship among the adjacent input frames. 
This is followed by two $5 \times 5$ and two $3 \times 3$ shaped convolutional kernels. The first $5 \times 5$ convolution is followed by a $2 \times 2$ max-pool operation to further reduce both the spectral and temporal resolution. 
Each of the convolution operations is followed by a $\mathtt{ReLU}$ activation and batch-normalization and dropout regularization (with dropout probability of $0.1$) layers. 
The LSTM layer consists of $64$ cells and is followed by a fully-connected layer with $1$ neuron (final prediction).
%
We employed $\mathtt{Adam}$ optimizer with a batch size of $256$ and an
initial learning rate of $0.001$, and a learning rate scheduler which reduces the learning rate by a factor of $0.1$ every $10$ epoch if there is no improvement in the validation metric. 
The $51300$ parameters ($205$~KB) of the model are trained up to $100$ epochs.
The complete model architecture is shown in Figure~\ref{fig:model-overview}.

\subsection{Baselines}
\label{sec:baselines}
We use two primary baselines for our experiments which are described below.

\myparagraph{MOS Prediction with Convolutional LSTM Backbone}
\label{sec:baseline-mos}
Our first baseline is the backbone model described in Section~\ref{sec:backbone}, where we train the model using the MOS ground truth only. Every other model proposed further in Section~\ref{sec:models} shares the same architecture, but simple modifications are made to accommodate the auxiliary labels and alternative loss functions.

\myparagraph{DNSMOS}
\label{sec:baseline-dnsmos}
The second baseline model is DNSMOS~\cite{reddy2021dnsmos}, a convolutional neural network based multi-stage self-teaching model inspired by continual lifelong learning~\cite{parisi2019continual}.
Our primary intention for including this model as a baseline is that of a sanity check as we note 
that comparing DNSMOS with the rest of the models proposed in this paper is not a fair comparison since (a) DNSMOS employs a more sophisticated multi-stage self-teaching architecture compared to our \textit{backbone} model, and (b) we employ 3.5x more audio samples in our training regimen.
Nevertheless, we use the same test set from the DNSMOS model to evaluate all proposed models. 

\subsection{Models Developed}
\label{sec:models}
We developed a number of models to incorporate the extra supervision (variance of the scores or histogram information) in addition to the MOS labels.
A high variance score is indicative of higher disagreement between the judges,  hence the variance ground truth can be a measurement of the confidence of the MOS scores. This confidence of the MOS scores can either be integrated as a weight to the loss function to give higher weight to the confident MOS scores (i.e., low variance) in a single task learning setup or it can be used directly as auxiliary ground truth in a multi-task learning setup. 
In the same vein, since there are only 5 possible values of the opinion scores (i.e., $1-5$), regardless of the number of opinion scores per sample, the ground truth of the opinion scores can be expressed as a 5-bin histogram and directly used to train the \textit{backbone} model. These approaches have the added flexibility of not requiring a fixed number and order of judges across the whole dataset, 
and are better suited for datasets collected with crowd-sourcing based approaches such as ITU-T P.808~\cite{reddy2020interspeech,naderi2020open}.

\subsubsection{Single Task MOS Estimation with Variance Weighted Loss}
\label{sec:single-task-variance}
We train the \textit{backbone} model with mini-batch gradient descent and loss is calculated for each sample in the batch before taking a mean across the batch to derive the final loss.
However, in this setup, we use the standard deviation ground truth to assign weight to each sample and calculate a weighted loss -- by assigning a higher weight to the samples with lower variance. This can be achieved in two primary ways:

\myparagraph{Inverse Variance Weighting}
This approach is inspired by~\cite{sinha2011statistical}, where the weight of each sample is calculated as $1/(\sigma_i + \delta)$ where $\sigma_i$ is the standard deviation of the sample and $\delta$ is a small constant (e.g., $10^{-3}$) to avoid division by zero.

\myparagraph{Linear Variance Weighting}
The numerical range of the opinion scores is $1-5$, and the range of the standard deviation is $0-2$. Inverse variance weighting can assign a high weight to samples with very low variance and as an alternative, we also explore the linear variance weighting strategy. Here samples with the highest $\sigma=2$ are assigned a weight of 0.1 and samples with the lowest $\sigma=0$ are assigned a weight of 1. And the weight of the remaining samples is linearly interpolated between the two extremes.

\subsubsection{Multi-Task Learning}
We experimented with several ideas on how-to incorporate extra supervision on the distribution of the opinion scores in a multi-task learning setup. They can be categorized as: (i) directly using the variance or median ground truth as the auxiliary label, (ii) calculating a 5 bin histogram of the opinion scores and using that as ground truth, and (iii) predicting opinion scores directly.

\myparagraph{MOS + Standard Deviation/Median Prediction}
In this setup, an extra regression head is added to the final layer of the backbone model that predicts the standard deviation or median of the opinion scores and is trained with the associated ground truth.

\myparagraph{Histogram Prediction}
The final layer of the backbone model predicts a 5 bin histogram of the opinion scores and is trained with the associated ground truth calculated from the individual opinion scores from the dataset. As the number of option scores per sample varies between 2 to 30 in our dataset, by creating a 5 bin histogram (to account for the 5 distinct values) we have a consistent way of representing the opinion distribution of all the samples. We experimented with 3 different loss functions to match the histogram distribution with the ground truth: (a) cross-entropy loss (b) Wasserstein loss~\cite{hou2017squared} (c) chi-square~\cite{pele2010quadratic,wang2021chi} loss. The MOS predictions can be derived by taking the weighted average of the bin values.

\begin{figure}[!htb]
    \centering
    \includegraphics[width=.95\linewidth]{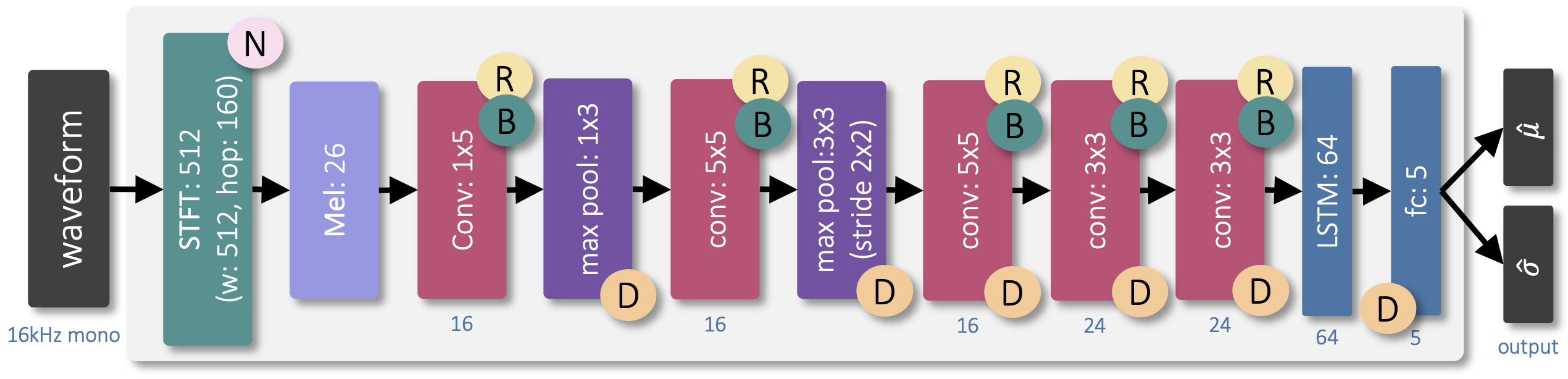}
    \caption{The model used in Direct Opinion Score Prediction.}
    \label{fig:model-vote-pred}
\end{figure}

\myparagraph{Direct Opinion Score Prediction}
In this setup (shown in Figure~\ref{fig:model-vote-pred}), we designate 5 neurons (since 75\% of the samples have 5 individual opinion scores) in the final layer of the backbone model as a representation of 5 judges and let them predict individual opinion scores. 
Since we have a variable number of opinion scores per sample and the real judges between the samples are not consistent (due to crowd-sourcing), it is not possible to directly compare the predicted and ground truth opinion scores to calculate the loss. Instead, we calculate MOS, standard deviation, median, etc. from the predicted opinion scores and calculate the losses against their respective ground truth from the samples. We experimented with two activation functions: (a) $\mathtt{ReLU}$, (b) Modified Sigmoid (i.e. $1 + 4 \times \mathtt{Sigmoid}(x)$) to predict values always between $1-5$ range.

\section{Evaluation Criteria}
\label{sec:exp}

We use (i) Pearson's correlation coefficient (PCC),
(ii) Spearman's rank correlation coefficient (SRCC) 
(iii) mean absolute error (MAE) and
(iv) root mean square error (RMSE) between the predicted MOS scores and the ground truth human ratings to evaluate the performance of our models. Since we are interested in evaluating the performance of a number of DNS models in enhancing the speech quality of the given samples, in addition to calculating the four evaluation metrics on a \textit{per-file} basis, we also group the clips together by the DNS model being used to generate them and calculate the evaluation metrics. This way of generating the evaluation metrics is referred to as \textit{stack-ranked} evaluation~\cite{reddy2021dnsmos}.

\section{Results}
\label{sec:results}
\begin{figure}[!htb]
    \centering
    \includegraphics[width=\columnwidth]{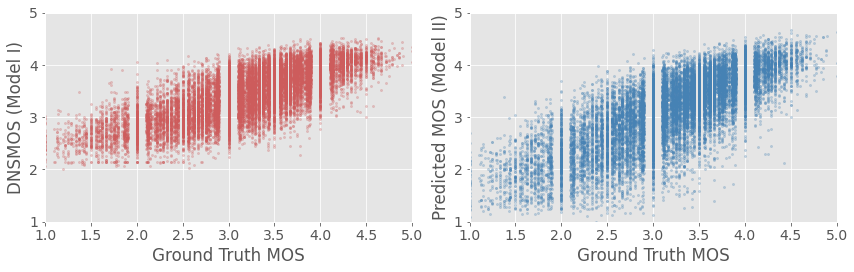}
    \caption{Scatter plot of real MOS labels (x-axis) vs predicted MOS labels (y-axis) for DNSMOS (model I) and 
    ConvLSTM Baseline (model II)}
    \label{fig:distro-dnsmos-vs-convlstm}
\end{figure}

\myparagraph{Baseline Sanity Check}
\label{sec:prelim}
\begin{table*}[!htb]
\centering
\caption{Comparison of Model Performance. Best metric per column marked in bold.}
\label{tab:results}
\resizebox{\textwidth}{!}{%
\begin{tabular}{@{}rlcllllllllll@{}}
\toprule
\multirow{2}{*}{ID} & \multirow{2}{*}{Model} & \multirow{2}{*}{Task} & \multicolumn{1}{c}{\multirow{2}{*}{\begin{tabular}[c]{@{}c@{}}Ground Truth\\ Labels\end{tabular}}} & \multicolumn{1}{c}{\multirow{2}{*}{Loss}} & \multicolumn{4}{c}{Per file} & \multicolumn{4}{c}{Stack Ranked} \\ \cmidrule(l){6-9}\cmidrule(l){10-13}  
 &  &  & \multicolumn{1}{c}{} & \multicolumn{1}{c}{} & \multicolumn{1}{c}{PCC} & \multicolumn{1}{c}{SRCC} & \multicolumn{1}{c}{MAE} & \multicolumn{1}{c}{RMSE} & \multicolumn{1}{c}{PCC} & \multicolumn{1}{c}{SRCC} & \multicolumn{1}{c}{MAE} & \multicolumn{1}{c}{RMSE} \\ \midrule
I & DNSMOS & Single & MOS & MSE & 0.7527 & 0.7473 & 0.4097 & 0.5247 & 0.9519 & 0.9514 & 0.2402 & 0.253 \\
II & ConvLSTM & Single & MOS & MSE & 0.7654 & 0.764 & 0.3735 & 0.4788 & 0.977 & 0.952 & 0.08507 & 0.1205 \\
III & ConvLSTM + Inverse Variance Weighting & Single & MOS, $\sigma$ & MSE & 0.7045 & 0.711 & 0.4325 & 0.5523 & 0.964 & 0.9137 & 0.0816 & 0.1058 \\
IV & ConvLSTM + Linear Variance Weighting & Single & MOS, $\sigma$ & MSE & 0.7656 & 0.7627 & 0.3803 & 0.4854 & 0.9788 & 0.956 & 0.09266 & 0.122 \\
V & ConvLSTM + Variance of Opinion Scores & Multi & MOS, $\sigma$ & MSE & 0.7669 & 0.762 & 0.3749 & 0.478 & \textbf{0.983} & \textbf{0.9597} & 0.07875 & 0.1131 \\
VI & ConvLSTM + Median of Opinion Scores & Multi & MOS, Median & MSE & 0.7718 & 0.7701 & 0.3652 & 0.4665 & 0.9777 & 0.9532 & 0.0775 & 0.1095 \\
VII & ConvLSTM + Histogram Prediction & Multi & Histogram & Cross Entropy & 0.766 & 0.7624 & 0.3712 & 0.4731 & 0.9725 & 0.9548 & 0.07309 & 0.1044 \\
VIII & ConvLSTM + Histogram Prediction & Multi & Histogram & Wasserstein & 0.7633 & 0.7647 & 0.3775 & 0.4826 & 0.9749 & 0.9528 & 0.08688 & 0.1265 \\
IX & ConvLSTM + Histogram Prediction & Multi & Histogram & Chi Square & 0.7576 & 0.7597 & 0.4025 & 0.5183 & 0.9739 & 0.9548 & 0.1457 & 0.1801 \\
X & ConvLSTM + Opinion Score (ReLU) & Multi & MOS, $\sigma$ & MSE & \textbf{0.7747} & \textbf{0.7742} & \textbf{0.3631} & \textbf{0.4635} & 0.9797 & 0.9529 & \textbf{0.07167} & \textbf{0.1043} \\
X1 & ConvLSTM + Opinion Score (Sigmoid) & Multi & MOS, $\sigma$ & MSE & 0.771 & 0.7689 & 0.3685 & 0.4709 & 0.9712 & 0.9456 & 0.08709 & 0.1169 \\ \bottomrule
\end{tabular}%
}
\end{table*}
The results of our ablation study are shown in Table~\ref{tab:results}. Our Convolutional LSTM based backbone (Model II), 
achieved similar stack ranked SRCC to DNSMOS (Model I) but shows 0.16 MAE and 0.13 RMSE improvement.
We perform a further inspection of the distribution of the predicted MOS labels generated by these two baselines against the ground truth, which is shown in Figure~\ref{fig:distro-dnsmos-vs-convlstm}.
The predictions of DNSMOS are heavily compressed between the 2-4.25 range (note Figure 2(d) of~\cite{reddy2021dnsmos}) while model II baseline predicts between a broader 1-4.7 range. The differences in model architecture (DNSMOS being more sophisticated) and training set size (model II using 2.5x samples) are the likely cause of such discrepancies, but it would require an in-depth investigation to find the concrete reasons.

\myparagraph{Effect of Auxiliary Supervision}
\label{sec:effect-aux}
Almost in every case, providing additional supervision 
leads to a better performance 
over our model II baseline.
Among our single task experiments,
where we employ the variance of the opinion scores to scale per sample loss term,
linear variance weighting strategy (IV) improves stack ranked SRCC by 0.4\% over model II, but inverse variance weighting (III) incurs a 3.83\% drop in the same metric.

\begin{table}[!htb]
\centering
\caption{Stack-ranked SRCC per bin for the three histogram prediction models.}
\label{tab:histigram-loss}
\resizebox{\columnwidth}{!}{%
\begin{tabular}{@{}rclllll@{}}
\toprule
\multicolumn{1}{l}{\multirow{2}{*}{ID}} & \multirow{2}{*}{Histogram Loss} & \multicolumn{5}{c}{Bins} \\ \cmidrule(l){3-7} 
\multicolumn{1}{l}{} &  & \multicolumn{1}{c}{1} & \multicolumn{1}{c}{2} & \multicolumn{1}{c}{3} & \multicolumn{1}{c}{4} & \multicolumn{1}{c}{5} \\ \midrule
VII & Cross Entropy & 0.9371 & 0.9351 & 0.5544 & 0.9464 & 0.9431 \\
VIII & Wasserstein & 0.9565 & 0.9548 & 0.631 & 0.9435 & 0.9149 \\
IX & Chi Square & 0.9355 & 0.9343 & 0.6758 & 0.948 & 0.9343 \\ \bottomrule
\end{tabular}%
}
\end{table}

\begin{figure}[!htb]
    \centering
    \includegraphics[width=\columnwidth]{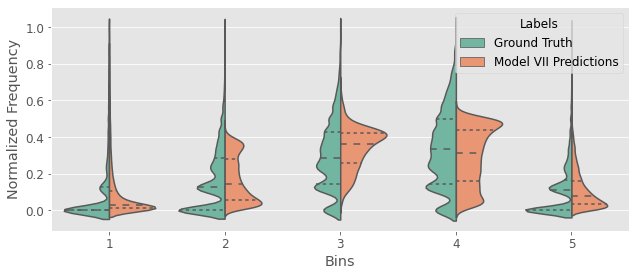}
    \caption{Violin plot~\cite{hintze1998violin} per histogram-bin for ground truth and model VII predictions.}
    \label{fig:his-pred-ce-gt-vs-pred.png}
\end{figure}

Among the three histogram prediction models, the cross-entropy (model VII) and chi-square loss (model IX) variants provide 0.28\% stack ranked SRCC improvement over the Model II baseline. We take a deeper look into them in Table~\ref{tab:histigram-loss}, where we notice that all three models struggle to predict the accurate probability of $\mathtt{Score}=3$ bin, indicated by much lower SRCC compared to other bins.
We further compare the ground truth and predictions for model VII in Figure~\ref{fig:his-pred-ce-gt-vs-pred.png} where we notice the model tends to learn a higher value (compared to ground truth) for $\mathtt{Score}=3$ bin.
%

\begin{figure}[!htb]
    \centering
    \includegraphics[width=\columnwidth]{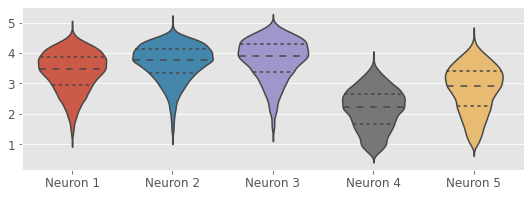}
    \caption{Violin plot~\cite{hintze1998violin} of the final 5 neurons'
    activations from the opinion score prediction (model X).}
    \label{fig:vote-pred-distro}
\end{figure}

According to the stack ranked PCC and SRCC metric, predicting MOS and variance score together (model V) results in the top performance improvement (0.66\% and 0.77\% respectively) compared to the model II baseline. In the rest of the 6 metrics, however, opinion score prediction with $\mathtt{ReLU}$ activation (model X) and MOS with median score prediction (Model VI) are the top two performing models.
Opinion score prediction with  $\mathtt{ReLU}$ activation (model X) achieved the highest improvement in RMSE (0.015 per-file, 0.016 stack-ranked) and SRCC (1.02\% per-file, 0.77\% stack-ranked).
To further investigate how model X generates the top results, we plot the distributions of the
activations from the final 5 neurons of model X in Figure~\ref{fig:vote-pred-distro}. We can notice that the first 3 neurons tend to produce higher scores than the last 2. The last two neurons also produce scores with relatively high variance.



\section{Conclusion}
\label{sec:conclusion}
In this paper, we demonstrated that deep neural network based mean opinion score (MOS) estimation of speech signals processed by DNS models can be improved by adding auxiliary supervision on the original distribution of the scores.
We demonstrated several ways these extra supervisions can be incorporated, either by integrating the uncertainty (variance of the scores) into a single task loss weighting strategy or directly incorporating the variance or histogram information into a multi-task learning setting. While some of the approaches appear to be more effective than others, it is clear that providing auxiliary supervision will result in better performance than doing single task MOS estimation. This benefit is practically 
free since during the data curation process (e.g., ITU-P.808~\cite{naderi2020open}) these statistics are typically available but discarded during model training. 
We also note that direct opinion score prediction seems to consistently generate the best results among all the proposed models. 

Our results were obtained with limited hyper-parameter search; our multi-task learning setups do not employ any loss balancing techniques~\cite{liu2019loss,kendall2018multi,ganin2014unsupervised} -- often crucial for achieving the best performance. 
We also opted for a simple convolutional LSTM model as our \textit{backbone} for the simplicity of exposition; combining auxiliary supervision into more sophisticated architectures (e.g. teacher-student model from DNSMOS) has the potential to bring substantial performance benefits. Further investigation is also warranted for a combination between the presented approaches.
It would be interesting to see whether the integration of higher-order moments (skewness, kurtosis) into the multi-task learning setup can induce further improvements.
We would also like to investigate the compatibility of our proposed approaches in more recent speech quality assessment challenges~\cite{dubey2022icassp} and datasets~\cite{reddy2021dnsmos835} where 
background noise quality labels are also being provided. In the same vein, we wish to also investigate the effect of  providing supervision in the form of soft labels regarding the reverberation of the speech signals (e.g. energy ratio C50~\cite{gamper2020blind}, reverberation time $T_{60}$~\cite{gamper2018blind}) in improving the quality of MOS estimation.

\section{Acknowledgments}
We would like to thank Sebastian Braun and the rest of the members of the Audio and Acoustics Research Group at Microsoft Research for their valuable feedback; Hari Dubey and Ross Cutler from the IC3-AI team for providing the dataset for the experiments.

\bibliographystyle{IEEEtran}
\bibliography{mybib}


\end{document}